\documentstyle[a4,12pt]{article}
\begin{document}
\begin{flushright}
LPT-ORSAY 99/87 \\
hep-th/9910219
\end{flushright}
\vskip 1cm
\begin{center}
{\Large \bf Brane cosmological evolution in a bulk with cosmological constant}\\
\vskip 2cm
{Pierre Bin\'etruy$^1$, C\'edric Deffayet$^1$, Ulrich Ellwanger$^1$, David Langlois$^2$\\ 
\vskip 1cm 
$^1$ LPT\footnote{Unit\'e mixte de recherche UMR n$^o$ 8627.}, 
Universit\'e Paris-XI, B\^atiment 210, 
F-91405 Orsay Cedex, France;\\
$^2$ D\'epartement d'Astrophysique Relativiste et de Cosmologie,
C.N.R.S.,\\ Observatoire de Paris, 92195, Meudon, France.} \\

\vskip 1.6cm 
{\bf Abstract.}  
\vskip 1cm
\end{center} 

{ We consider the cosmology of a ``3-brane universe'' in a five dimensional (bulk) space-time with a
cosmological constant. We show that Einstein's equations admit a first integral, analogous to the first
Friedmann equation, which governs the evolution of the metric in the brane, 
whatever the time evolution 
of the metric along the fifth dimension.
We thus obtain the cosmological evolution in the brane for 
{\it any equation of state} describing 
the matter in the brane,
 without needing  the dependence of the metric on the fifth dimension.
In the particular case  $p = w \rho$, $(w = constant)$, we give explicit expressions 
for the time evolution of the brane scale factor, which show that standard cosmological evolution
can be obtained (after an early non conventional phase) in a scenario \`a la Randall and Sundrum, where a
brane tension compensates the
bulk cosmological constant.
We also show that a tiny deviation from exact compensation leads to an effective cosmological
constant at late time.
Moreover, when the  metric along the fifth dimension is static, 
we are able to extend the solution
found on the brane to the whole spacetime.}

\def\beq{\begin{equation}}
\def\eeq{\end{equation}}
\def\d{\delta}
\def\fourG{{{}^{(4)}G}}
\def\4R{{{}^{(4)}R}}
\def\H{{\cal H}}
\def\K5{{\kappa}}
\def\K52{{\kappa^2}}
\def\C{{\cal C}}
\def\lamb{{\rho_{\Lambda}}}

\newcommand{\A}{A}
\newcommand{\B}{B}
\newcommand{\mmu}{\mu}
\newcommand{\mnu}{\nu}
\newcommand{\ii}{i}
\newcommand{\jj}{j}
\newcommand{\jl}{[}
\newcommand{\jr}{]}
\newcommand{\ml}{\sharp}
\newcommand{\mr}{\sharp}

\newcommand{\da}{\dot{a}}
\newcommand{\db}{\dot{b}}
\newcommand{\dn}{\dot{n}}
\newcommand{\dda}{\ddot{a}}
\newcommand{\ddb}{\ddot{b}}
\newcommand{\ddn}{\ddot{n}}
\newcommand{\pa}{a^{\prime}}
\newcommand{\pb}{b^{\prime}}
\newcommand{\pn}{n^{\prime}}
\newcommand{\ppa}{a^{\prime \prime}}
\newcommand{\ppb}{b^{\prime \prime}}
\newcommand{\ppn}{n^{\prime \prime}}
\newcommand{\fda}{\frac{\da}{a}}
\newcommand{\fdb}{\frac{\db}{b}}
\newcommand{\fdn}{\frac{\dn}{n}}
\newcommand{\fdda}{\frac{\dda}{a}}
\newcommand{\fddb}{\frac{\ddb}{b}}
\newcommand{\fddn}{\frac{\ddn}{n}}
\newcommand{\fpa}{\frac{\pa}{a}}
\newcommand{\fpb}{\frac{\pb}{b}}
\newcommand{\fpn}{\frac{\pn}{n}}
\newcommand{\fppa}{\frac{\ppa}{a}}
\newcommand{\fppb}{\frac{\ppb}{b}}
\newcommand{\fppn}{\frac{\ppn}{n}}

\newcommand{\dA}{\dot{A_0}}
\newcommand{\dB}{\dot{B_0}}
\newcommand{\fdA}{\frac{\dA}{A_0}}
\newcommand{\fdB}{\frac{\dB}{B_0}}

\section{Introduction}

There has been recently a lot of activity on the possibility that we live 
in a three-dimensional world embedded in a higher dimensional space. 
Modifying the old Kaluza-Klein picture \cite{bl87}, where the extra-dimensions must 
be sufficiently compact, these recent developpements are based on 
 the idea that ordinary matter fields could be confined to a 
three-dimensional world,  corresponding to our apparent Universe, 
while gravity could live in a higher dimensional space \cite{add98}. The usual constraints
on Kaluza-Klein models could therefore be relaxed \cite{add98b}, and ``large'' 
extra-dimensions would be conceivable, thus leading to a fundamental 
Planck mass much lower than its apparent three-dimensional value, even 
as low as the $TeV$ scale. In the same spirit, 
Gogberashvili \cite{gog} and 
 Randall and Sundrum \cite{rs99a},\cite{rs99b}, reviving an older idea
 \cite{old}, have recently pointed out that the 
extra dimension (single in their case) need not even be compact.

A first important question concerning these kinds of models is to check 
whether they give back standard four-dimensional gravity in the brane assumed 
to represent our world. When the gravitational field generated by the 
brane itself is taken into account, it is not so easy to check that usual 
gravity is recovered. Randall and Sundrum showed that this is the 
case because of the existence of massless gravitons  trapped in 
the brane, although small corrections are expected from  
 massive modes due to the presence of a fifth dimension. 
Several groups have also begun to work on possible  experimental 
signatures of these kinds of models {\cite{exp}.

In addition to the study of the weak field limit, one would  expect 
that cosmology as well could 
provide   constraints on these types of models. An interesting aspect of 
cosmology in such models is that, thanks to the 
symmetries, it is possible to deal with the non-linearities of the Einstein's 
equations and to detect effects that may not appear in a linearized 
treatment.
The first crucial question is the viability of 
such models with respect to  the cosmological evolution of our Universe. 
In a previous 
paper \cite{bdl99}, three of us  have shown  that, generically, the equations governing 
the cosmological evolution of the brane will be different from the 
analogous Friedmann equations of standard cosmology (the same holds for a brane with finite 
thickness \cite{kkop}). Essentially, the difference 
lies in the fact that the energy density of the brane appears quadratically 
in the new Friedmann equations in contrast with the linear behaviour of 
the usual equations. In the simplest model where the bulk is supposed to 
be empty (and stabilized), we found explicit solutions which showed that the 
corresponding cosmological scenario would be incompatible 
with nucleosynthesis constraints. 

Following our work, it was noticed  by two groups (\cite{cgkt99},\cite{cgs99})
that a possible way to reconcile 
brane cosmology with the required standard cosmological scenario (at least 
since nucleosynthesis) would be to use Randall and Sundrum's approach 
\cite{rs99b}, i.e. 
to put a (negative) cosmological constant in the bulk  
and to introduce in the brane, in addition  to ordinary matter, 
 a constant tension that would exactly compensate the bulk cosmological 
constant so that the, quadratic, leading order due to the brane contribution 
would be cancelled, leaving the next, linear, order.

These works however only provided an approximate perturbative analysis 
of the cosmological evolution. Now, exact solutions of the global 
Einstein's equations are extremely useful, and until now, such solutions with a bulk
cosmological constant exist only  
when the spacetime is static (\cite{rs99a}), or in the extensions given by
 Kaloper \cite{Kaloper99} (see also \cite{cr99} and 
\cite{japan}), 
where the brane tension 
does not exactly compensate the bulk cosmological constant, so that 
dynamical evolution is possible, but in the simplest form of exponential 
expansion. Some other exact solutions are also known in the context 
of general supergravity \cite{cs97} or Horava-Witten supergravity 
\cite{low}.  General conditions 
for static universes can be found in \cite{Ellwanger99}.

The purpose of the present work is to solve  the five-dimensional 
Einstein's equations for {\it any type of matter in the brane} with a cosmological
 constant in the bulk. We first show in section 2 that Einstein's equations admit a first integral,
  which in particular provides directly the cosmological evolution of the brane. In the following 
  section, we show that an additional assumption, namely that the metric along the fifth dimension does
  not evolve in time, enables us to solve for the whole space-time metric, i.e. to find explicitly the
  dependence of the metric on the transverse coordinate as well as time. Finally, in the last section, we
  solve analytically, for different cases, the new Friedmann equation obtained  in the present
  work and discuss the consequences.

\section{Solving Einstein's equations}

Let us present here the general framework.    We shall consider  
five-dimensional spacetime metrics of the form 
\begin{equation}
ds^{2}=\tilde{g}_{AB}dx^A dx^B = 
g_{\mmu \mnu} dx^{\mmu}dx^{\mnu} + b^{2}dy^{2}, 
\end{equation}
where $y$ is the coordinate of the fifth dimension.
Throughout this article, we will focus our attention on 
the hypersurface defined by $y=0$, which we  identify
with the  world volume of the brane that  forms our universe.
 Since we are interested in cosmological solutions, we take a metric of the form
\begin{equation}
ds^{2}=-n^{2}(\tau,y) d\tau^{2}+a^{2}(\tau,y)\gamma_{ij}dx^{i}dx^{j}+b^{2}(\tau,y)dy^{2},
\label{metric}
\end{equation}
where $\gamma_{ij}$ is a maximally symmetric 3-dimensional metric ($k=-1,0,1$ will 
parametrize the spatial curvature).

The five-dimensional Einstein equations take the usual form 
  \beq
{\tilde G}_{\A\B}\equiv{\tilde R}_{\A\B}-{1\over 2}{\tilde R}{\tilde g}_{\A\B}
  =\kappa^2 \tilde{T}_{\A\B}, \label{einstein}
\eeq
where ${\tilde R}_{\A\B}$ is the five-dimensional Ricci tensor, 
${\tilde R}={\tilde g}^{\A\B}{\tilde R}_{\A\B}$ the scalar curvature and 
the constant $\kappa$  is  related to
the  five-dimensional Newton's constant $G_{(5)}$ and the 
 five-dimensional reduced Planck mass ${M}_{(5)}$,
by the relations
\beq
 \kappa^2 = 8 \pi G_{(5)} = {M}_{(5)}^{-3}.
\eeq

Inserting the ansatz (\ref{metric}) for the metric, the non-vanishing 
components of the Einstein tensor 
${\tilde G}_{\A\B}$ are found to be 
\begin{eqnarray}
{\tilde G}_{00} &=& 3\left\{ \fda \left( \fda+ \fdb \right) - \frac{n^2}{b^2} 
\left(\fppa + \fpa \left( \fpa - \fpb \right) \right) + k \frac{n^2}{a^2} \right\}, 
\label{ein00} \\
 {\tilde G}_{\ii\jj} &=& 
\frac{a^2}{b^2} \gamma_{ij}\left\{\fpa
\left(\fpa+2\fpn\right)-\fpb\left(\fpn+2\fpa\right)
+2\fppa+\fppn\right\} 
\nonumber \\
& &+\frac{a^2}{n^2} \gamma_{ij} \left\{ \fda \left(-\fda+2\fdn\right)-2\fdda
+ \fdb \left(-2\fda + \fdn \right) - \fddb \right\} -k \gamma_{ij},
\label{einij} \\
{\tilde G}_{05} &=&  3\left(\fpn \fda + \fpa \fdb - \frac{\dot{a}^{\prime}}{a}
 \right),
\label{ein05} \\
{\tilde G}_{55} &=& 3\left\{ \fpa \left(\fpa+\fpn \right) - \frac{b^2}{n^2} 
\left(\fda \left(\fda-\fdn \right) + \fdda\right) - k \frac{b^2}{a^2}\right\}.
\label{ein55} 
\end{eqnarray} 
In the above expressions, a prime stands for a derivative with respect to
 $y$, and a 
dot for a derivative with respect to $\tau$. 

The stress-energy-momentum tensor can be decomposed into two parts, 
\begin{equation}
\tilde{T}^\A_{\quad \B}  =  \check{T}^\A_{\quad \B}|_{_{\rm bulk}} 
+ T^\A_{\quad \B}|_{_{\rm brane}},
\end{equation}
where $\check{T}^\A_{\quad \B}|_{_{\rm bulk}}$ is the energy momentum tensor 
of the bulk matter, which will be assumed in the present work to be of the form
\begin{equation}
\check{T}^\A_{\quad \B}|_{_{\rm bulk}}= \mbox{diag} 
\left(-\rho_B,P_B,P_B,P_B,P_T \right), 
\label{bulksour}
\end{equation}
where the energy density $\rho_B$ and pressures $P_B$ and $P_T$ are independent of the 
coordinate y. Later, we will be specially interested in the case of a cosmological constant 
for which $-\rho_B=P_B=P_T$. 

The second 
term $T^\A_{\quad \B}|_{_{\rm brane}}$ corresponds to the matter content
in the 
brane $(y=0)$. Since we consider here only strictly homogeneous and isotropic 
geometries inside the brane, 
the latter can be expressed quite generally in the form
\begin{equation}
T^\A_{\quad \B}|_{_{\rm brane}}= \frac{\delta (y)}{b} \mbox{diag} 
\left(-\rho_b,p_b,p_b,p_b,0 \right), 
\label{source}
\end{equation}
where the energy density $\rho_b$ and pressure $p_b$ are independent of the 
position inside the brane, i.e. are functions only of time.
We may also add some other brane sources with similar energy momentum tensor.

The assumption that $\tilde{T}_{05}=0$, which physically means that there is no flow of matter
along the fifth  dimension, implies that ${\tilde G}_{05}$ vanishes. It 
 then turns out, remarkably,  that 
 the components $(0,0)$ and $(5,5)$ of Einstein's  equations 
(\ref{einstein}) (with
  (\ref{ein00}) and (\ref{ein55})) in the bulk, can 
be rewritten in the simple form
\begin {eqnarray}
F^\prime &=& \frac{2 a^\prime a^3}{3}\K52\check{T}^0_{\quad 0} \label{cons1},\\
\dot{F} &=& \frac{2 \dot{a} a^3}{3}\K52\check{T}^5_{\quad 5} \label{cons2},
\end{eqnarray}
where $F$ is a function of $\tau$ and $y$ defined by
\begin{equation}
F(\tau,y)\equiv \frac{\left(a^\prime  a\right)^2}{b^2} -\frac{\left(\dot{a}  a\right)^2}{n^2}-ka^2.
 \end{equation}

Since $\check{T}^0_{\quad 0}=-\rho_B$ is here independent of $y$, one can integrate (\ref{cons1}),
 which gives
\beq
F+ {\K52 \over 6} a^4\rho_B+\C=0, \label{firstint}
\eeq
where $\C$ is a constant of integration which a priori depends on time. 
Assuming in addition that $\check{T}^0_{\quad 0}=\check{T}^5_{\quad 5}$, one finds using the time
derivative of (\ref{cons1}) and the $y-$derivative of (\ref{cons2}) that $\rho_B$ is constant in time.
This also implies that $\C$ is constant in time. 
In order to deal with the  last component of Einstein's equations 
(see (\ref{einij})),
 it is convenient to use the
Bianchi identity,
\beq
\nabla_A \tilde{G}^{A0} = 0,
\eeq
which can be rewritten (using ${\tilde G}_{05}=0$, and $\tilde{T}^0_{\quad 0}=\tilde{T}^5_{\quad 5}$) in the form 
\beq
\partial_{\tau}\left(\frac{F^\prime}{a^\prime}\right)=\frac{2}{3}\dot{a}a^2\gamma_{j}^i
\tilde{G}^{j}_i.
\eeq
One finds (using (\ref{cons1})) that this equation is identically satisfied if $-\rho_B=P_B$.
Hence, when the bulk source is a cosmological constant, any set of functions $a$, $n$, and $b$ 
satisfying (\ref{firstint}) or, more explicitly, 
\begin{equation}
\left(\frac{\dot{a}}{na}\right)^2=\frac{1}{6}\K52\rho_B+
\left(\frac{a^\prime}{ba}\right)^2 -\frac{k}{a^2}+
\frac{\C}{a^4},
\label{integral}
\end{equation} 
together with
${\tilde G}_{05}=0$,  will be solution of
all Einstein's equations (\ref{einstein}), locally in the bulk.

The brane can then be  taken into account by using the junction conditions \cite{Israel66}, which 
simply relate the jumps of the derivative of the metric across 
 the brane to the stress-energy tensor inside the brane.
This procedure is described in detail in our previous paper \cite{bdl99} 
in the context 
of five-dimensional brane cosmology (see also \cite{cr99}). 
The relevant expressions are 
\begin{eqnarray}
\frac{\jl a^\prime \jr}{a_0 b_0}&=&-\frac{\K52}{3}\rho_b, \label{aarho}\\
\frac{\jl n^\prime \jr}{n_0 b_0}&=&\frac{\K52}{3}  \left( 3p_b + 2\rho_b \right), 
\label{nnrho}
\end{eqnarray}
where the subscript $0$ for $a,b,n$ means that these functions are 
taken in $y=0$, and $\jl Q \jr = Q(0^+)-Q(0^-)$ denotes the jump of the function $Q$ across $y = 0$.

Assuming the symmetry $y \leftrightarrow -y$ for simplicity, the junction condition 
(\ref{aarho}) can be used to compute $a^\prime$ on the two sides of the brane, 
and by continuity when $y \rightarrow 0$,  (\ref{integral}) will yield 
the  generalized (first) Friedmann equation (after setting $n_0=1$
 by a suitable change of time):  
\beq
{\dot a_0^2\over a_0^2}={ \K52 \over 6}\rho_B+{ \kappa^4 \over 36}\rho_b^2+{\C\over a_0^4}
- {k \over a_0^2}.
 \label{friedfried}
\eeq
The salient features of this equation are that, first, the bulk energy 
density enters linearly, second, the brane energy density enters quadratically, 
and finally the cosmological evolution depends on a free parameter $\C$ (related
to the choice of initial conditions in the whole space-time), whose 
influence corresponds to an effective radiation term (from the standard 
point of view, i.e. interpreting linearly the additional term)\footnote{From a technical 
point of view, the presence of a free parameter, in contrast 
to the standard cosmological equations, can be explained in the following 
way. In  four-dimensional cosmology, the component (0,0) of Einstein's equations
yields a first integral, because it contains only first derivatives in time. 
In the present work, we have used the component (0,0) to integrate in $y$, 
whereas the time integration is operated with the 
help of the  component (5,5) which contains second derivatives in time.}.
 One can note that this equation
 is a generalization of the exact equation obtained in our previous paper \cite{bdl99}. 
{\it This equation is enough to study the cosmological evolution in the  brane, independently of the metric
outside and in particular of the time evolution of $b$}. The analysis of this equation will be postponed
until section 4. In the next section we will be interested in obtaining an 
explicit solution for the whole five dimensional metric.

\section{Explicit dependence on the fifth dimension  for a  stabilized bulk}
In this section, it will be shown that, with the help of an additional assumption, namely that 
 the fifth dimension is static, in the sense that 
\beq
\dot b=0, \label{bstatic}
\eeq
it is then
possible to solve the full space-time metric, i.e. to determine the 
explicit dependence of the metric on the coordinate $y$.
The restriction (\ref{bstatic}) allows us to go to the gauge 
\beq
b=1.
\eeq
It then follows immediately from 
equation $\tilde G_{05}=0$,  that 
$n$ can be expressed in terms of $a$ according to the relation
\beq
{\dot a\over n}=\alpha(t), \label{alpha}
\eeq
where $\alpha$ is a function that depends only on time (and not on $y$).
Inserting this into (\ref{cons1}) yields the following differential equation:
\beq
\alpha^2+k-(aa')'={\K52 \over 3}\rho_B a^2,
\eeq
which is valid everywhere in the bulk (but not in the brane)
 on the two sides of the brane separately. 
It can be integrated  in $y$, yielding 
\beq
a^2=A\cosh(\mu y)+B \sinh(\mu y) +C , \label{a2y}
\eeq
with 
\beq
\mu=\sqrt{-{2 \K52 \over 3}\rho_B}, 
\eeq
in the case $\rho_B <0$, or
\beq
a^2=A\cos(\mu y)+B \sin(\mu y) +C , 
\eeq
with 
\beq
\mu=\sqrt{{2 \K52 \over 3}\rho_B}, 
\eeq
in the case $\rho_B >0$,
or finally 
\beq
a^2=\left(\alpha^2+k\right) y^2+Dy+E,
\eeq
for $\rho_B=0$.
In the following, we will focus on the first case $\rho_B <0$, but all the 
equations will apply as well to the case $\rho_B >0$, up to the transformation
$\mu \rightarrow i\mu$, $B\rightarrow -i B$. 

The coefficients $A$, $B$, $C$, $D$, $E$ are functions of time, and $C$ is 
expressible in terms of $\alpha$ as  
\beq
C=3 \frac{\alpha^2+k}{\K52 \rho_B}, \label{defC}
\eeq
and the others can be determined by the junction conditions.
The symmetry $y \leftrightarrow -y$ imposes the relations 
 $A_+ =A_-\equiv \bar{A}$, $B_+ =-B_-\equiv \bar{B}$ between the  coefficients on the two 
 sides of the brane. Using (\ref{aarho}) and (\ref{nnrho}),  one then  finds 
\beq
{\mu \bar{B}\over \bar{A}+C}=-{\K52 \over 3}\rho_b,\quad 
{2 \mu \dot {\bar{B}}\over \dot {\bar{A}}+\dot C}=\K52 \left( p_b+{1\over 3}\rho_b \right). 
\label{junction}
\eeq
Note that one can check explicitly energy conservation in the brane from these relations, 
i.e. 
\beq
\dot \rho_b+3{\dot a_0\over a_0}(\rho_b+p_b)=0. \label{cons}
\eeq

Let us finally fix our temporal gauge, by imposing the condition $n_0=1$, i.e. 
we choose as time what corresponds to the cosmic time in the brane.
The function $\alpha$ is then simply $\dot a_0$ (see equation (\ref{alpha})), and (\ref{a2y}) 
specialized to $y=0$ gives
\beq
\bar{A}=a_0^2-C,
\eeq
with, from equations (\ref{defC}) and (\ref{friedfried}),
\beq
C={a_0^2 \over 2} \left(1+{\K52 \rho_b^2 \over 6 \rho_B} \right) + {3 \C \over \K52 \rho_B a_0^2}.
\eeq
Using the first relation in (\ref{junction}), the remaining coefficient 
is determined as 
\beq
\bar{B}=-{\K52 \rho_b\over 3\mu}a_0^2.
\eeq
Substituting back the coefficients $A$ and $B$ thus obtained in the general expression (\ref{a2y})
and using (\ref{friedfried}), one finally obtains 
the following expression for 
the  scale factor $a(t,y)$:
\begin{eqnarray}
a(t,y)&=&\left\{{1\over 2}\left(1+{\K52 \rho_b^2\over 6\rho_B}\right) a_0^2 
+{3\C  \over \K52\rho_B a_0^2} \right. \cr &&
+ \left[ {1\over 2}\left(1-{\K52 \rho_b^2\over 6\rho_B}\right) a_0^2 
-{3\C  \over \K52\rho_B a_0^2}\right]\cosh (\mu y) \cr &&
\left. 
-{\kappa \rho_b\over\sqrt{ -6\rho_B}}a_0^2 \sinh(\mu |y|)\right\}^{1/2}.
\end{eqnarray}
The other coefficient of the metric, $n(t,y)$, follows directly 
from the above expression 
with the help of  the relation (\ref{alpha}), i.e. 
\beq
n(t,y)={\dot a(t,y)\over \dot a_0(t)}.
\eeq
Therefore, from the two functions $a_0(t)$ and $\rho_b(t)$ {\it restricted to the brane}, 
which are obtainable
by solving the coupled system constituted of the two equations (\ref{friedfried}) and (\ref{cons}), 
one is able to infer the {\it extension of the metric in the bulk away from the brane}. 

Let us mention briefly the case $\rho_B=0$. Following the same procedure as above, one would end up
with the following expression for the scale factor:
\beq
a(t,y)=\left\{a_0^2-{\K52 \rho_b  \over 3}a_0^2 |y|+\left\{{\kappa^4 \rho_b^2\over 36} a_0^2 
+{\C\over  a_0^2}\right\}y^2\right\}^{1/2}.
\eeq
In the special case $\C=0$, one recognizes the linear solutions obtained in our previous paper 
\cite{bdl99}.

We have not specified here whether the fifth dimension is compact or not.
If it is compact, then one must check that one gets in the end a globally 
well defined solution (see \cite{bdl99}).

The next section will be devoted to the determination of the evolution of the scale factor 
on the brane, $a_0(t)$, using equations (\ref{friedfried}) and (\ref{cons}), 
which - we wish to emphasize - does not depend on the hypothesis (\ref{bstatic}) of the present
section.

\section{Cosmological scenarios}
In this section we will examine the consequences of equation (\ref{friedfried}) on 
the cosmological evolution inside the brane. In general, 
 one still finds  the quadratic behaviour 
in the brane energy density which leads to a non conventional cosmology \cite{bdl99}. 
However, as suggested by \cite{cgkt99} and \cite{cgs99},
we will show that  one can indeed implement the idea of Randall and Sundrum
\cite{rs99b} in the cosmological context to recover standard cosmology.
We will find exact solutions that match the early non
conventional cosmology to the subsequent ordinary cosmology. 

Let us thus  assume that the energy density in 
the brane can be decomposed into two parts,
\beq
\rho_b=\lamb +\rho, 
\eeq
where $\lamb$ is a constant that represents an intrinsic tension of the brane
 and $\rho$ stands for the  ordinary energy density 
in cosmology. Substituting in (\ref{friedfried}) one gets
\beq
{\dot a_0^2 \over a_0^2}={\K52 \over 6} \rho_B+{\kappa^4 \over 36}\lamb^2 +
{\kappa^4 \over 18}\lamb \rho 
+{\kappa^4 \over 36}\rho ^2+{\C\over a_0^4} -{k \over a_0^2}. \label{rsfried}
\eeq
If we follow Randall and Sundrum by choosing $\lamb$ such that 
\beq
{\K52 \over 6}\rho_B+{\kappa^4 \over 36}\lamb^2=0,
\eeq
then one sees that  standard cosmology is recovered with the identification 
(\cite{cgkt99}, \cite{cgs99})
\beq
8 \pi G \simeq {\kappa^4 \lamb \over 6},
\eeq
when $\rho \ll \lamb$. 
Let us however keep $\lamb$ unspecified at this stage. In the case $\rho=0$ and $\C=0$
 one would recover the solutions given by Kaloper \cite{Kaloper99}.

To get analytic solutions we will now  assume that ordinary matter is described
 by an equation of state 
of the form $p=w\rho$, with $w$ constant. Then, using equation (\ref{cons}) 
for  $\rho$ and $p$, 
one can write 
\beq
\rho=\rho_* (a_0/a_*)^{-q}, \quad q=3(1+w),
\eeq
where $\rho_*$ and $a_*$ are constants.
Inserting this expression in (\ref{rsfried}),  it is possible to integrate explicitly the resulting 
equation in the case where $\C=0$ and $k=0$. 
Assuming the first term on the right hand side of (\ref{rsfried}) to be  positive and 
defining
\beq
\lambda=\sqrt{{ \rho_B\over 6 \K52}+{\lamb^2\over 36}},
\eeq
one finds, for $\lambda>0$, 
\beq
a_0=a_*\rho_*^{1/q}\left\{ {\lamb\over 36\lambda^2}\left[\cosh\left(q \K52 \lambda t\right)-1\right]
+{1\over 6\lambda} \sinh\left(q \K52 \lambda t\right)\right\}^{1/q},
\eeq
and, for $\lambda=0$, 
\beq
a_0(t)=a_*\left(\kappa^2\rho_*\right)^{1/q}\left({q^2 \over 72} 
\kappa^2\lamb  t^2+{q\over 6}t\right)^{1/q}
\eeq
(the origin of time being chosen so that $a_0(0)=0$).
It is clear, from the latter expression, in the case $\lambda=0$,
 how one passes from a very early universe, 
characterized by a non-conventional evolution $a(t)\sim t^{1/q}$, to a late time phase described 
by standard cosmology, $a(t)\sim t^{2/q}$. In the case $\lambda>0$, with $\lambda$ 
sufficiently small, one  obtains three successive phases, a non-conventional phase dominated 
by $\rho^2$, a conventional phase dominated by $\rho$ and, finally, an exponential phase, where 
$\lambda$ plays the r\^ole of an effective cosmological constant in our Universe. 
Therefore, brane cosmology is able to produce a late time acceleration phase such
as seems to be needed from the latest cosmological observations (\cite{supernovae}), at the price 
however of a tiny mismatch in the compensation between $\rho_B$ and the tension of the 
brane. The required  fine-tuning  expresses the brane cosmological 
version of the well-known cosmological constant problem.

Let us finally examine  the case of a universe filled with radiation ($w=1/3$).
It is then possible, still with $k=0$,  to integrate explicitly the case 
$\C \neq 0$, because the $\C$-term has the same dependence on $a_0$ 
as the term proportional to $\rho$ in (\ref{rsfried}). Moreover, 
the free parameter $\C$ can be constrained by nucleosynthesis. 
Indeed, at the time of nucleosynthesis, the universe is dominated by 
the radiation energy density, 
which can be written  
\beq
\rho_{rad}(t_N)=g_*  \frac{\pi^2}{30} T_N^4,
\eeq
where $g_*$ is the effective number of relativistic degrees of freedom at
 that time. In the standard model, $g_*(standard)=10.75$, and any deviation 
$\Delta g_*$ is 
strongly constrained by the observed abundances of light elements, 
typically $\Delta g_* < 2$.  
In our model, since the additional $\C$-term evolves like radiation, it 
can also be seen  effectively as  additional relativistic degrees of freedom, 
subject to the usual constraint, so that 
\beq  
\rho_\C(t_N)\equiv {3\C\over 8 \pi G a^4(t_N)}\leq \frac{\pi^2}{15} T_N^4
\eeq
at nucleosynthesis. 

To conclude, we wish to recall the main results of this work. First, 
it has been shown that one can obtain a first integral of Einstein's 
equations, which provides, on the brane, a relation  analogous to the 
(first) Friedmann
equation and which depends only on the geometry and matter content {\it of 
the brane},  except for a constant parameter. Second, when $\dot b=0$,  one 
can extend explicitly the solution found on the brane to the whole spacetime.
Finally, we have shown, exhibiting  exact solutions, that brane 
cosmology appears compatible, at late times, with standard cosmology, 
in a Randall-Sundrum type model.  
In this respect, an important question for future investigations 
is wether an (exact or quasi)
Randall-Sundrum configuration can be naturally reached in a dynamical scenario
for the very early universe.

\vskip 1cm

Note added:
Several papers which discuss related issues have appeared quasi simultaneously 
to this one
\cite{sms99,flanagan99} (see also \cite{other}).

\end{document}